\documentstyle[epsfig]{elsart} 
\input{epsf} 
 
\begin{document} 
 
%%%%{\hskip 4in	} 

%%{\hskip 4in	paper  draft} 
 
%%{\hskip 4in	December/26/02} 

%%%%%%\vskip 2in 

\def\ss{\scriptscriptstyle} 
\def\ysub{_{\scriptscriptstyle Y}} 
\def\ybarsub{_{\scriptscriptstyle \overline{Y}}} 
\def\ysubb{_{\scriptscriptstyle Y - \scriptscriptstyle \overline{Y}}}
\def\xf{x_{\scriptscriptstyle F}} 
\def\pt{p_{\scriptscriptstyle T}^2} 
\def\pbar{\overline{p}} 
\def\Do{D^{0}}  
\def\Dobar{\overline{D}^{0}}
\def\Ds{D^+_{\mbox{\footnotesize{s}}}}
\def\Lm{\Lambda^{\, 0}}  
\def\Lbar{\overline{\Lambda}^{\, 0}}
\def\Lc{\Lambda_c^{\, +}}
\def\Lcbar{\Lambda_c^{\, -}}
\def\Xim{\Xi^-}
\def\Xibar{\Xi^+} 
\def\Om{\Omega^-} 
\def\Ombar{\Omega^+}
\def\PYTHIA{\textsc{Py\-thia}} 
\def\JETSET{\textsc{Jet\-set}} 
 
\def\etal{{\em et al.}} 
\def\PL#1{Phys. Lett. {#1}} 
\def\PR#1{Phys. Rev. {\bf#1}} 
\def\PRL#1{Phys. Rev. Lett. {\bf#1}} 
\def\NP#1{Nucl. Phys. {\bf#1}} 
\def\REV#1{Annu. Rev. Nucl. Part. Sci. {\bf#1}}
\def\NIM#1{Nucl. Instr. and Meth. {#1}}
%%%%\begin{center} 
%%%%{\Large Asymmetries in the Production of $\Lm$ 
%%%%in 250 GeV/$c$ $\pi^\pm$, K$^\pm$ and p -- Nucleon Interactions} 
 
%%%%\noindent Target Journal: 
%%%%\vskip 2in 
%%%%\end{center} 
 
\newpage 
\begin{frontmatter} 
                       
\title{Asymmetries in the Production of $\Lm$ 
in 250 GeV/$c$ $\pi^\pm$, K$^\pm$ and p -- Nucleon Interactions} 
%\collab{Fermilab E769 Collaboration} 
%%%%\input{author.tex} 
%%
%\documentstyle [12pt]{article}
%\textwidth=17 true cm
%%%\evensidemargin=1.5 true cm  
%\begin{document}
\renewcommand{\thefootnote}{\alph{footnote}}
\vskip 0.5 true cm
%\centerline{\LARGE \bf  A Colabora\c c\~ao E769}  
%\vspace{1 cm}
\begin{center}{\normalsize}
G.A.~Alves,$^{(1)}$
S.~Amato,$^{(1),}${\footnote {Now at Universidade Federal do Rio de Janeiro, Rio
de Janeiro, Brazil}}
J.C.~Anjos,$^{(1)}$
J.A.~Appel,$^{(2)}$
J.~Astorga,$^{(5)}$
T.~Bernard,$^{(5),}${\footnote {Deceased}
S.B.~Bracker,$^{(4),}${\footnote {Retired}}
L.M.~Cremaldi,$^{(3)}$
W.D.~Dagenhart,$^{(5)}${\footnote {Now at Brandeis University, Waltham, MA 02454}}           
C.L.~Darling,$^{(8)}$
D.~Errede,$^{(7),}${\footnote {Now at University of Illinois, Urbana, Il 61801}}
H.C.~Fenker,$^{(2),}${\footnote {Now at Thomas Jefferson National Accelerator 
Facility,  Newport News, VA 23606}} 
C.~Gay,$^{(4),}${\footnote {Now at Yale University, New Haven, CT 06511}}
D.R.~Green,$^{(2)}$                 
R.~Jedicke,$^{(4),}${\footnote {Now at LPL, University of Arizona, Tucson, AZ
85721}}
P.E.~Karchin,$^{(6)}$
S.~Kwan,$^{(2)}$                 
L.H.~Lueking,$^{(2)}$
%P.M.~Mantsch,$^{(2)}$
J.R.T.~de~Mello~Neto,$^{(1),~{\mathtt a}}$
J.~Metheny,$^{(5)}$                 
R.H.~Milburn,$^{(5)}$
J.M.~de~Miranda,$^{(1)}$
H.~da~Motta,$^{(1)}$
A.~Napier,$^{(5)}$                   
M.S.~Nicola,$^{(1)}$
D.~Passmore,$^{(5)}$
A.~Rafatian,$^{(3)}$
A.C.~dos~Reis,$^{(1)}$
W.R.~Ross,$^{(8)}$
A.F.S.~Santoro,$^{(1),}${\footnote {Now at Universidade do Estado do Rio de
Janeiro, Rio de Janeiro, Brazil}}
M.~Sheaff,$^{(7)}$                      
M.H.G.~Souza,$^{(1)}$
C.~Stoughton,$^{(2)}$
M.E.~Streetman,$^{(2),~{\mathtt c}}$                  
D.J.~Summers,$^{(3)}$
S.F.~Takach,$^{(6)}$
%L.~Chen-Tokarek,$^{(2)}$
A.~Wallace,$^{(8)}$
Z.~Wu$^{(8)}$ 
\vskip 0.5 true cm
$^{(1)}$Centro Brasileiro de Pesquisas F\'\i sicas, Rio de Janeiro, Brazil  \\
$^{(2)}$Fermi National Accelerator Laboratory, Batavia, Illinois, 60510 \\
$^{(3)}$University of Mississippi, University, MS  38677 \\
$^{(4)}$University of Toronto, Toronto, Ontario, Canada, M5S 1A7 \\
$^{(5)}$Tufts University, Medford, MA 02155 \\
$^{(6)}$Wayne State University, Detroit, Michigan 48202 \\
$^{(7)}$University of Wisconsin, Madison, WI  53706 \\
$^{(8)}$Yale University, New Haven, CT 06511 \\
}
\vskip 0.5 true cm
{\bf(Fermilab E769 Collaboration)} 
\end{center} 
 
%\end{document}

%%\addtolength{\topmargin}{-.5in}
%%\addtolength{\textwidth}{.5in}
%%\setlength{\textheight}{8.5in}
\vfill \eject
\begin{abstract} 
Using data from Fermilab fixed-target experiment E769, we have measured
particle-antiparticle production asymmetries for $\Lm$ 
hyperons in 250 GeV/$c$ $\pi^\pm$, K$^\pm$ and
p -- nucleon interactions.
The asymmetries are measured as 
functions of Feynman-x ($\xf$) and $\pt$ over the ranges 
$-0.12\leq\xf\leq 0.12$ and  $0\leq\pt\leq 3 (GeV/c)^2$ (for positive beam) and 
$-0.12\leq\xf\leq 0.4$ and $0\leq\pt\leq 10 (GeV/c)^2$ (for negative beam).  We find 
substantial asymmetries, even at $\xf$ around zero. We also observe
leading-particle-type asymmetries. These latter effects are qualitatively 
as expected from valence-quark content of the target and variety of 
projectiles studied.

\end{abstract} 
\end{frontmatter}

Leading particle production effects have been studied both
experimentally~\cite{javier,charm-exp,charm-expb,charm-expc,sudeshna,strange-modl} and
theoretically~\cite{charm-mod,charm-modb,charm-modc,charm-modd,charm-mode}. 
These effects are manifest as an enhancement in
the production rate of particles that share one or more valence quarks with
an initial state hadron compared to their antiparticles when they
share either none or fewer.  This enhancement is expected to become
larger as the produced particles carry more and more of the initial
particle's center-of-mass momentum. Other effects, like the associated 
production of a kaon and a hyperon, can also contribute to an 
asymmetry in hyperon-antihyperon production~\cite{capella1}.

        As a byproduct of
our charm program in Fermilab Experiment E769, we 
 collected a large sample
of $\Lm$ and $\Lbar$ 
 hyperons which we have used
to measure the particle-antiparticle production 
 asymmetries reported here. Given E769's variety of identified beam particles, 
 we can study the production asymmetries as the content of valence quarks 
 in the beam changes. This is the first such measurement in the central
 kinematical region (i.e., near $\xf=0$) and with this variety of 
 projectiles in a single experiment.

The asymmetry $A$ can be defined as 
%%%%% 

%\begin{equation} 
%A \equiv {\frac {N_{\Lm} - r_{\Lm}~N_{\Lbar} }{N_{\Lm} + r_{\Lm}~N_{\Lbar}}} \;
%~;~~ {r_{\Lm} = {\epsilon_{\Lm} \over \epsilon_{\Lbar}}}
%\label{eq1}
%\end{equation} 

\begin{equation} 
A \equiv {\frac {N_{\Lm} - N_{\Lbar} }{N_{\Lm} + N_{\Lbar}}} \;
\label{eq1}
\end{equation}

%%%%%                    
where $N_{\Lm}$ ($N_{\Lbar}$) is the number of $\Lm$ ($\Lbar$) 
produced over the kinematic range of interest.

 For all beam types, a positive asymmetry growing larger with increasingly 
negative $\xf$  is expected, because $\Lm$ ($uds$) shares two valence 
 quarks with the target p ($uud$) or n ($udd$) 
 while $\Lbar$ ($\overline{u}\overline{d}\overline{s}$) shares none. For 
 both the $\pi^+$ ($u\overline{d}$) and $\pi^-$ ($\overline{u}d$) beams, an 
asymmetry close to zero is expected for $\xf>0$,  because $\Lm$ 
and $\Lbar$ each share one valence 
quark with the incident beam. For the K$^+$ ($u\overline{s}$) beam, 
a negative asymmetry which becomes 
more negative with increasing $\xf$ is expected for $\xf>0$, 
because $\Lbar$ shares the 
heaviest valence quark with the beam while $\Lm$ shares only a 
light quark.  The situation is reversed for the K$^-$ ($\overline{u}s$) 
beam, where the asymmetry in the forward direction is expected to be
positive and to become more positive with increasing $\xf$. For the 
p ($uud$) beam, a large positive asymmetry which grows even larger 
with increasing $\xf$ is expected in the forward direction, because 
$\Lm$ shares two valence quarks with beam proton 
while $\Lbar$ shares none. Although measurements of $\Lm$ and $\Lbar$ 
production asymmetries have
been made in several other experiments 
\cite{javier,accmor1,add-evid,add-evidb,add-evidc,add-evidd}, no single
experiment to date has been able to study the asymmetries and to compare
them for all five beam types, $\pi^-$, $\pi^+$, $p$, $K^+$ and $K^-$.

 Experiment E769 recorded about $400 \times 10^6$ physics events from
 interactions of 250 GeV/$c$ hadron beams of both signs on a multifoil 
 target of Be, Cu, Al and W. The
 negative beam consisted of $93\%$ $\pi^-$, $5\%$ $K^-$ and $1.5\%$ $p^-$, the
 positive beam of $61\%$ $\pi^+$, $4.4\%$ $K^+$, and $34\%$ $p$. Event-by-event 
 beam particle identification was accomplished through the use of a differential
 \v{C}erenkov counter\cite{DISC} and a transition radiation 
 detector\cite{TRD}. Pre-scaling of triggers was used to enhance the 
 recorded sample of minority-beam-particle interactions.
 
 The apparatus in Fermilab experiment E769 has been previously described (See
 \cite{jeff,colin,E769} and references therein.) In this analysis we use the 11 silicon
 microstrip planes (1 - 30 cm downstream of the target), 35 drift chambers (150 -
 1750 cm downstream of the target), 2 multiwire proportional chambers (130 cm,
 180 cm downstream of the target), and 2 magnets (centered at 290 cm, 620 cm 
 downstream of
 the target) for momentum measurement. The two threshold \v{C}erenkov counters downstream of
 the target were not used and the electromagnetic and hadronic calorimeters  were
 used only for on-line event selection. The trigger required that the total
``transverse energy'' (i.e., sum of the products of energy
observed times the tangent of the angle from the target to each calorimeter
segment)
%absolute spatial first moment of the energy) observed in our downstream
%calorimeters
be at least 5.5 GeV. Most $\Lm$'s decay before entering the drift chamber
region, but downstream of the end of the
silicon vertex detectors. 
%The transverse-energy trigger was based on the energy deposited in the calorimeters
%and was highly efficient for charm and strange particle production events. 
 
%\noindent{\bf Event Selection} 

Throughout this paper, references to a particle should be taken to 
include its antiparticle except where explicitly stated otherwise. 
For historical reasons and current availability of data samples, the 
positive beam and negative beam samples come from rather different 
event selections.  The positive beam $\Lm$'s were reconstructed 
from $\Lm$'s decaying upstream of the silicon vertex detector, while the 
negative beam sample comes from $\Lm$'s decaying downstream and 
reconstructed using tracks which were not seen in the silicon vertex 
detector.  In addition to the resulting differences in selection 
criteria, there are more $\Lm$ data from the negative beam and 
consequently larger kinematic ranges are accessible.

All $\Lm$'s were reconstructed using the $p\pi^-$ decay mode.  Selection 
criteria were chosen to maximize the signal significance using Monte 
Carlo simulation for the signal, and using side-band data for background 
projections.  The ratio of proton to pion decay momenta was required to 
be larger than 3.0.  For the positive beams (p, $\pi^+$ and $K^+$), 
proton and $\pi^-$ tracks were required to make a vertex downstream of 
the last target, but upstream of the silicon detector, and have a distance 
of closest approach less than 0.02 cm.  The resulting $\Lm$ candidate 
track was then required to have an impact parameter relative to the 
interaction point of less than 0.006 cm.  For the negative beams 
($\pi^-$ and $K^-$), the proton and $\pi^-$ tracks were required to 
make a vertex downstream of the silicon detector, but upstream of 
the first analysis magnet.  The tracks had to have a distance of 
closest approach less than 0.7 cm at the decay vertex, with no 
requirement on the impact parameter of the lambda candidate track.

\begin{figure}[htb]
\centerline{\epsfig{figure=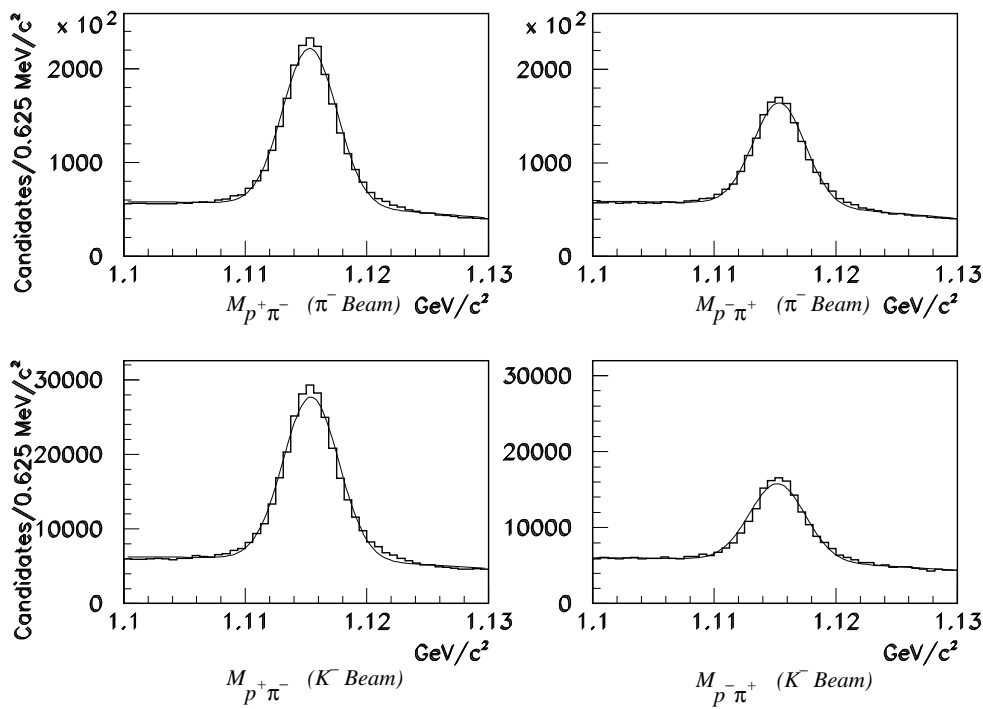,height=9.3cm}}
\protect\caption{Effective $p\pi$ mass distributions for the negative beams.}
\label{signalsneg}
\end{figure}

\begin{figure}[htb]
\centerline{\epsfig{figure=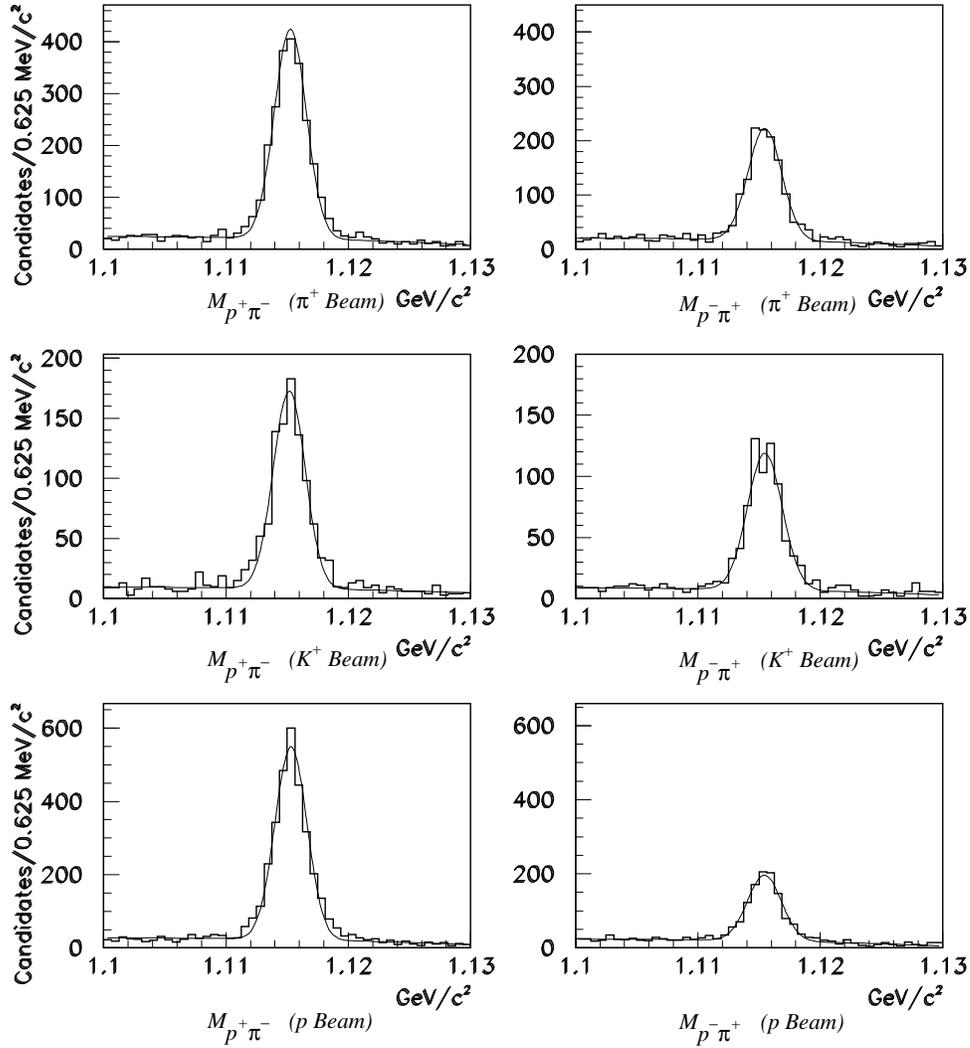,height=14cm}}
\protect\caption{Effective $p\pi$ mass distributions for the positive beams.}
\label{signalspos}
\end{figure}

The reconstructed mass 
distributions for each interval of $\xf$ and $\pt$ were  
fit in the mass range from $1.101$ to $1.127$ $GeV/c^2$ using a binned maximum likelihood method with a Gaussian signal plus a linear 
background. All $\xf$ and $\pt$ bin 
widths are much larger than the resolution of the variable binned.
%binning parameter.
In the fit, the central reconstructed mass values and mass resolutions were
fixed to values obtained from Monte Carlo simulation.
%%for each interval of $\xf$ and $\pt$ to be the same as those obtained
%%by fitting reconstructed E791 Monte Carlo data.
%generating a sample of Monte 
Monte Carlo simulation studies demonstrated that $K^0_s\rightarrow\pi^+\pi^-$ 
provided a flat background and produced 
a negligible effect on the fit
numbers of $\Lm$'s. The total reconstructed mass distributions are shown 
in Figs. \ref{signalsneg} and \ref{signalspos}. Table \ref{tablesig} shows the
total numbers of $\Lm$ and $\Lbar$ from the fits shown.

\begin{table*} [htb] \centering
\begin{tabular} {|c|l|l|} \hline
 Beam    &    $~~~~~~~~~~~\Lm$   &   $~~~~~~~~~~\Lbar$  \\ \hline
$\pi^+$  & ~~$~~~~1,965~\pm~49~~~$ & ~$~~1,053~\pm~37~~~$   \\ \hline
$\pi^-$  & $1,537,000~\pm~1,877$ & $996,200~\pm~1,712$ \\ \hline
K$^+$    & ~~$~~~~~~845~\pm~32~~~$ & ~~$~~~~621~\pm~28~~~$   \\ \hline
K$^-$    & $~~203,800~\pm~660~~$ & $~95,300~\pm~548~~$  \\ \hline
p        & ~$~~~~2,615~\pm~56~~~$ & ~~$~~~~919~\pm~35~~~$   \\ \hline
\end{tabular}
\protect\caption{Total number of $\Lm$ and $\Lbar$ observed for each beam.
The errors are statistical as determined by the fits.}
\label{tablesig}
\end{table*}

%\noindent{\bf Measurement of Asymmetry} 

\begin{figure}[b]
\centerline{\epsfig{figure=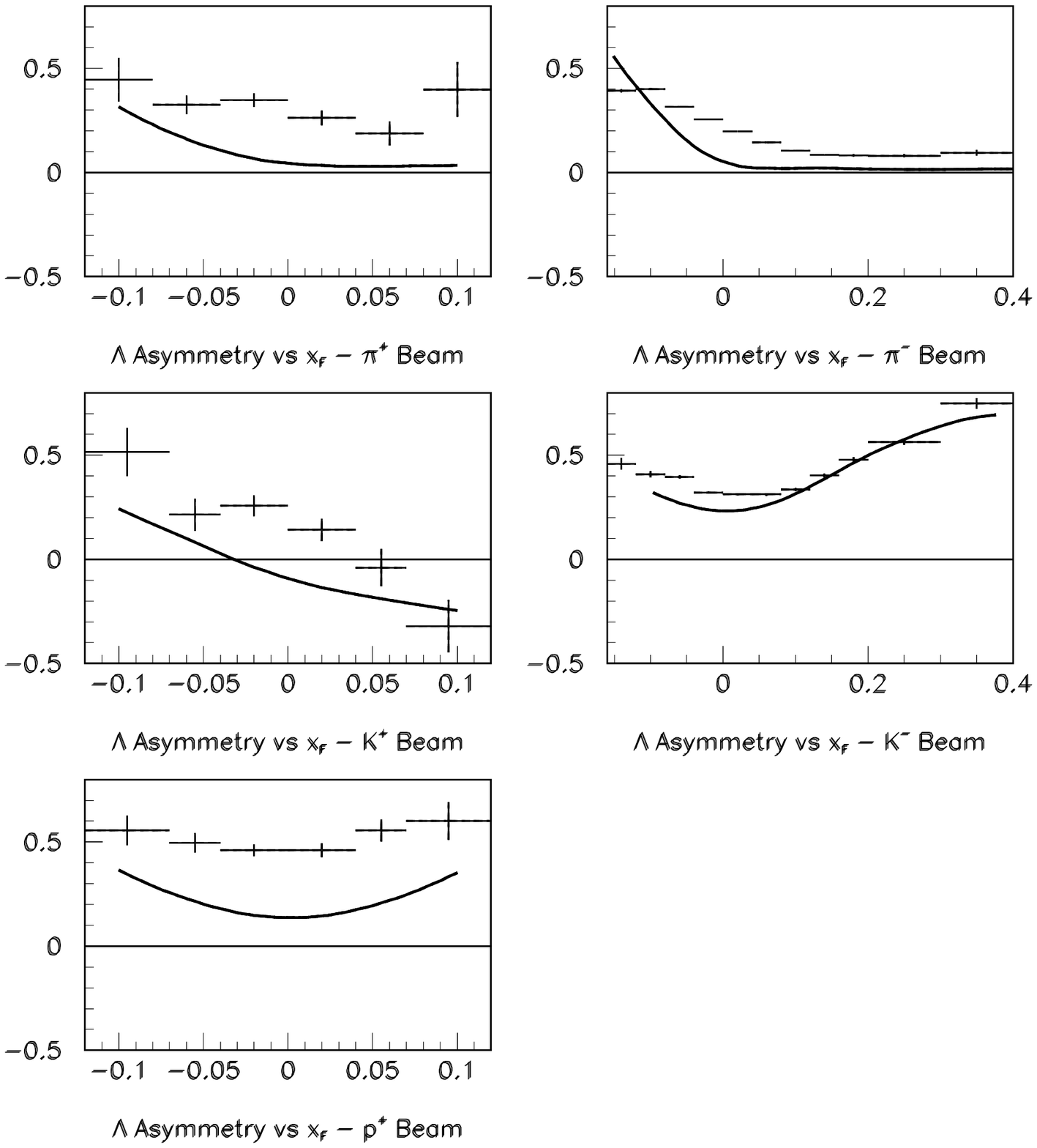,height=14cm}}
\protect\caption{$\Lm$ production asymmetries vs $x_F$ for various incident beams. The errors shown are statistical, 
with systematic errors being negligible. The curves are predictions of the \PYTHIA ~6.2 model.}
\label{ass_xf}
\end{figure}

\begin{figure}[b]
\centerline{\epsfig{figure=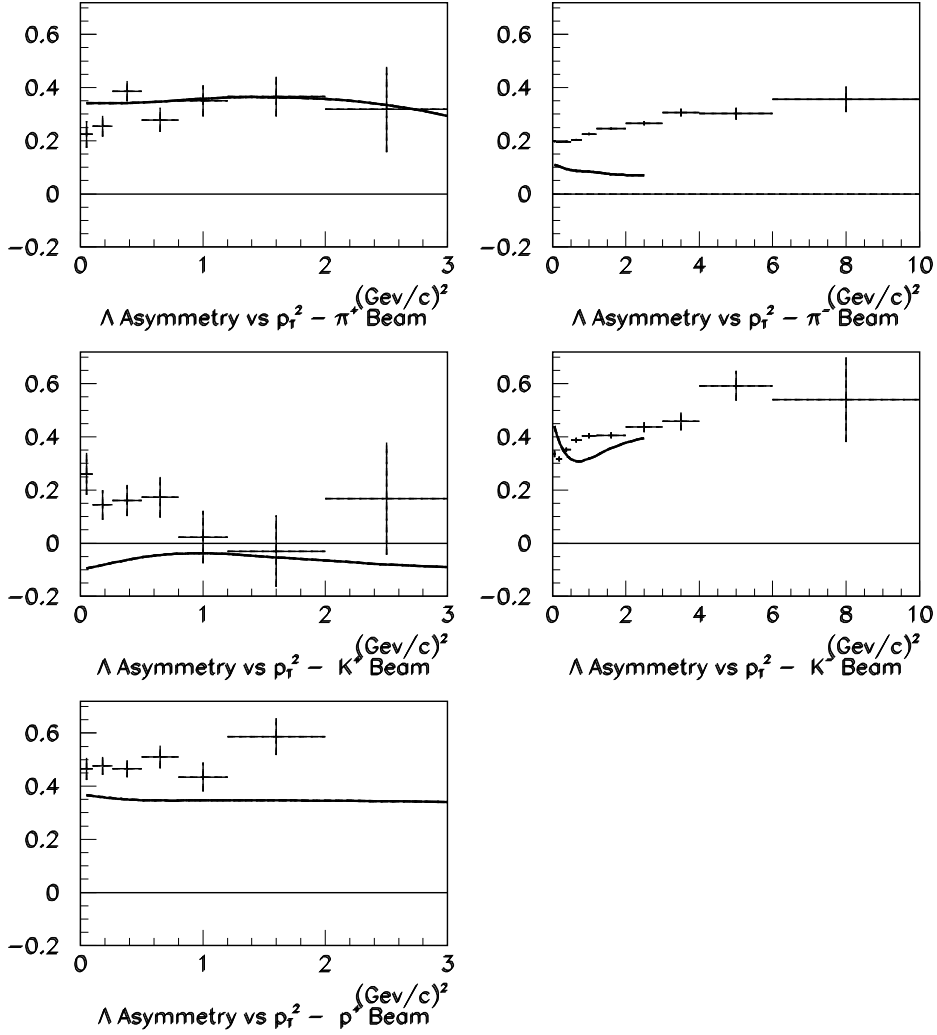,height=14cm}}
\protect\caption{$\Lm$ production asymmetries vs $\pt$ for various 
incident beams. The errors shown are statistical, 
with systematic errors being negligible. The curves are predictions of 
the \PYTHIA ~6.2 model. The prediction of \PYTHIA ~is limited to the range of 
$\pt$ shown.}
\label{ass_pt}
\end{figure}

      For each $\xf$ and $\pt$ bin we compute an 
asymmetry, $A$, as defined in Equation \ref{eq1}. Values for the $N$'s 
were obtained  
from individual fits to the $p \pi$ mass plots for events selected to 
lie  within each $\xf$ and $\pt$ interval. For each beam type, the asymmetry $A(\xf)$ integrated over $\pt$, and the asymmetry $A(\pt)$
integrated over our range of $\xf$, are presented in Figs.~\ref{ass_xf} and \ref{ass_pt}. 
%where a comparison of the three hyperon asymmetries can be easily made. 
The results are also listed in Tables~\ref{tabsysxfpos}, \ref{tabsysptpos},
\ref{tabsysxfneg}, and \ref{tabsysptneg} along with statistical errors.
 
The experimental apparatus itself could create an apparent particle-antiparticle
asymmetry if there were a difference in the detection efficiencies for $\Lm$ and
$\Lbar$. Selection criteria for the particle and antiparticle samples were identical. 
However, geometrical acceptances and reconstruction efficiencies were not
necessarily the same. To evaluate this potential effect, a large 
sample of simulated events was created using the \PYTHIA/\JETSET~event 
generators \cite{pyth,pythb}. These were passed through a detailed simulation of 
the E769 spectrometer to simulate ``data'' in digitized format, which was then
processed through the same computer reconstruction code as that used for data
from the experiment. Simulated event data were subjected to the same selection
criteria as used for detector data. No difference was found between the acceptances and
efficiencies for $\Lm$ compared to $\Lbar$ at a level significant with 
respect to the statistical errors. 

%uncertainties are quoted. 
 
%\noindent{\bf Sources of systematic uncertainties.} 
 
We also looked for systematic effects from the other following sources: 
\begin{itemize} 
\item {Event selection criteria;} 
\item {
The minimum transverse energy in the calorimeters required  
in the on-line event selection;} 
\item {K$_S$ contamination of the $\Lm$ signal;}
\item {Misidentification of beam particle types;}
\item {$\Lm$ from higher-mass hyperon decays;}
%\item {Uncertainties in calculating relative efficiencies for 
%particle and antiparticle;}
\item {Mass fits (shape)}; 
%\item {beam contamination;}

\end{itemize} 
 
All seven sources, the relative efficiencies and the six others listed, 
were found to produce systematic effects that were 
negligible in each kinematic bin, both individually and in 
the ensemble,  when compared to the statistical uncertainties. 

%Small errors came 
%from K$^-$ contaminating the $\pi^-$ beam. The results were then corrected 
%for this contamination.  

%These uncertainties are included as
%systematic errors in Tables~\ref{tabsysxfpos},\ref{tabsysptpos},
%\ref{tabsysxfneg}, and \ref{tabsysptneg}. 
 
%No change in the central values of asymmetry have been made for possible
%effects of beam contamination by kaons. 
%\noindent{\bf Discussion and Summary} 

The behavior of the asymmetries shown in Figs.~\ref{ass_xf} and \ref{ass_pt} 
have leading particle effects where expected. A larger asymmetry 
is observed when there is a larger difference in the
number of valence quarks in the $\Lm$ or $\Lbar$ in common with the target or
the various beams. We~\cite{E769b} and others~\cite{charm-exp,wa82} have reported 
evidence for similar effects in the production of $D^{\pm}$
mesons in the forward region ($\xf>0$). 
%where the leading particle effects appear
%in the production of charm hadrons, they do so at higher $\xf$.

   The \PYTHIA ~{\bf 6.2} ~\cite{pyth62} model describes only some features of
our results, and those only qualitatively, as can be seen in
Fig.~\ref{ass_xf} and Fig.~\ref{ass_pt}. This model predicts small  values of asymmetry for
$\xf=0$ in contrast to our results which have large asymmetries 
in this region. This may be due to associated production of strange mesons and
baryons (more $\Lm \overline{K}$ than $\Lbar K$). Leading particle  effects play an
increasingly important role as $\left|\xf\right|$ increases.    

Our  results for particle-antiparticle asymmetries are
consistent with the results obtained by other
experiments where similar data exists~\cite{javier,accmor1,accmor2}. Our results can be described 
qualitatively in terms of the  
energy thresholds for the production of hyperons and antihyperons  
together with their associated particles and a model in which  
the recombination of valence quarks in the beam and target  
particles contributes to the hyperon and antihyperon production in  
an asymmetrical manner~\cite{strange-mod}.

In summary, we report a precise, systematic study of the 
production asymmetries for $\Lambda$ hyperons by various incident hadrons 
in a single experiment. The range of $\xf$ covered allows
the study of asymmetries in regions close to $\xf=0$. Our results are
consistent with other experiments where similar data exists and 
while models so far describe some features of our results, they 
do so only qualitatively.
 
%\begin{ack}  
	We gratefully acknowledge the assistance of the staffs of Fermilab and 
of all the participating institutions. This research was supported by the 
Brazilian Conselho Nacional de Desenvolvimento Cient\'{\i}fico e Tecnol\'ogico,
 the U.S. 
National Science Foundation, the U.S. Department of Energy, and the National
Research Council of Canada. Fermilab is
operated by the Universities Research Association, Inc., under  contract with
the United States Department of Energy. 
%\end{ack} 

\newpage
% 
%\begin{figure}[b]
%\centerline{\epsfig{figure=eff_gimp.ps,height=14cm}}
%\protect\caption{The ratios of efficiencies 
%$\epsilon_{\Lm}/\epsilon_{\Lbar}$ as functions of $x_F$ and $\pt$}
%\label{eff_gimp}
%\end{figure}

\newpage 
% 
%%%\input tabres1.tex
%%%\input tabres2.tex
%%%\input tabres3.tex
%%%\input tabres4.tex
%%%
\begin{table} [ht] \centering
\begin{tabular} {|c|c|c|} \hline 
  Beam	&$x_F$ Region& Production Asymmetry              \\ \hline\hline

	&$-0.12~\leq~x_F~<~-0.08$    & $0.446~\pm~0.105$  \\ \cline{2-3} 
	&$-0.08~\leq~x_F~<~-0.04$    & $0.325~\pm~0.045$  \\ \cline{2-3}
	&$-0.04~\leq~x_F~<~~0.00$     & $0.349~\pm~0.033$  \\ \cline{2-3}
$\pi^+$	&$~0.00~\leq~x_F~<~+0.04$    & $0.263~\pm~0.036$  \\ \cline{2-3}
	&$+0.04~\leq~x_F~<~+0.08$    & $0.188~\pm~0.058$  \\ \cline{2-3}
	&$+0.08~\leq~x_F~\leq~+0.12$ & $0.399~\pm~0.132$  \\ \hline\hline

	&$-0.12~\leq~x_F~<~-0.08$  & $0.516~\pm~0.116$  \\ \cline{2-3} 
	&$-0.08~\leq~x_F~<~-0.04$  & $0.214~\pm~0.076$  \\ \cline{2-3}
	&$-0.04~\leq~x_F~<~~0.00$  & $0.258~\pm~0.050$  \\ \cline{2-3}
K$^+$	&$ 0.00~\leq~x_F~<~+0.04$  & $0.142~\pm~0.054$  \\ \cline{2-3}
	&$+0.04~\leq~x_F~<~+0.08$  & $-0.040~\pm~0.090$  \\ \cline{2-3}
	&$+0.08~\leq~x_F~\leq~+0.12$  & $-0.321~\pm~0.125$  \\ \hline\hline
	
	&$-0.12~\leq~x_F~<~-0.08$  & $0.556~\pm~0.071$  \\ \cline{2-3} 
	&$-0.08~\leq~x_F~<~-0.04$  & $0.497~\pm~0.046$  \\ \cline{2-3}
	&$-0.04~\leq~x_F~<~~0.00$  & $0.461~\pm~0.027$  \\ \cline{2-3}
p$^+$	&$ 0.00~\leq~x_F~<~+0.04$  & $0.461~\pm~0.032$  \\ \cline{2-3}
	&$+0.04~\leq~x_F~<~+0.08$  & $0.556~\pm~0.053$  \\ \cline{2-3}
	&$+0.08~\leq~x_F~\leq~+0.12$  & $0.601~\pm~0.089$  \\ \hline
\end{tabular}
\protect\caption{$\Lm$ production asymmetries vs $x_F$ for positive beams. 
The errors shown are statistical, with systematic errors being negligible.}
\label{tabsysxfpos}
\end{table}
%%%
\begin{table} [ht] \centering
\begin{tabular} {|c|c|c|} \hline 
Beam	&${p_T}^2$ Region& Production Asymmetry        \\ \hline\hline

	&$0.00 <{{p_T}^2}\leq 0.10$ & $~0.224~\pm~0.051$\\ \cline{2-3}
	&$0.10 <{{p_T}^2}\leq 0.26$ & $~0.255~\pm~0.039$\\ \cline{2-3}
	&$0.26 <{{p_T}^2}\leq 0.50$ & $~0.387~\pm~0.039$\\ \cline{2-3}
$\pi^+$	&$0.50 <{{p_T}^2}\leq 0.80$ & $~0.279~\pm~0.046$\\ \cline{2-3}
	&$0.80 <{{p_T}^2}\leq 1.20$ & $~0.350~\pm~0.059$\\ \cline{2-3}
	&$1.20 <{{p_T}^2}\leq 2.00$ & $~0.367~\pm~0.074$\\ \cline{2-3}
%	&$2.00 <{{p_T}^2}\leq 3.00$ & $~0.318~\pm~0.161$\\ \cline{2-3}
	&$2.00 <{{p_T}^2}\leq 3.00$ & $~0.318~\pm~0.161$\\ \hline\hline
%	&$3.00 <{{p_T}^2}\leq 4.00$ & $-0.015~\pm~0.334$\\ \hline\hline

	&$0.00 <{{p_T}^2}\leq 0.10$ & $~0.261~\pm~0.079$\\ \cline{2-3}
	&$0.10 <{{p_T}^2}\leq 0.26$ & $~0.144~\pm~0.056$\\ \cline{2-3}
	&$0.26 <{{p_T}^2}\leq 0.50$ & $~0.160~\pm~0.059$\\ \cline{2-3}
	&$0.50 <{{p_T}^2}\leq 0.80$ & $~0.173~\pm~0.077$\\ \cline{2-3}
K$^+$	&$0.80 <{{p_T}^2}\leq 1.20$ & $~0.023~\pm~0.099$\\ \cline{2-3}
	&$1.20 <{{p_T}^2}\leq 2.00$ & $-0.031~\pm~0.134$\\ \cline{2-3}
%	&$2.00 <{{p_T}^2}\leq 3.00$ & $~0.168~\pm~0.211$\\ \cline{2-3}
	&$2.00 <{{p_T}^2}\leq 3.00$ & $~0.168~\pm~0.211$\\ \hline\hline
%	&$3.00 <{{p_T}^2}\leq 4.00$ & $~0.002~\pm~0.498$\\ \hline\hline

	&$0.00 <{{p_T}^2}\leq 0.10$ & $~0.465~\pm~0.042$\\ \cline{2-3}
	&$0.10 <{{p_T}^2}\leq 0.26$ & $~0.476~\pm~0.034$\\ \cline{2-3}
	&$0.26 <{{p_T}^2}\leq 0.50$ & $~0.466~\pm~0.033$\\ \cline{2-3}
	&$0.50 <{{p_T}^2}\leq 0.80$ & $~0.509~\pm~0.043$\\ \cline{2-3}
p$^+$	&$0.80 <{{p_T}^2}\leq 1.20$ & $~0.434~\pm~0.054$\\ \cline{2-3}
	&$1.20 <{{p_T}^2}\leq 2.00$ & $~0.586~\pm~0.069$\\ \cline{2-3}
%	&$2.00 <{{p_T}^2}\leq 3.00$ & $~0.603~\pm~0.144$\\ \cline{2-3}
	&$2.00 <{{p_T}^2}\leq 3.00$ & $~0.603~\pm~0.144$\\ \hline\hline
%	&$3.00 <{{p_T}^2}\leq 4.00$ & $~0.236~\pm~0.309$\\ \hline\hline
\end{tabular}
\protect\caption{$\Lm$ production asymmetries vs ${p_T}^2$ for positive beams. 
The errors shown are statistical, with systematic errors being negligible.}
\label{tabsysptpos}
\end{table}
%%%
\begin{table} [ht] \centering
\begin{tabular} {|c|c|c|} \hline 
Beam	&$x_F$ region& Production Asymmetry    \\ \hline\hline

	&$-0.16~\leq~x_F~<~-0.12$  & $0.393~\pm~0.009$  \\ \cline{2-3}
	&$-0.12~\leq~x_F~<~-0.08$  & $0.400~\pm~0.005$  \\ \cline{2-3}
	&$-0.08~\leq~x_F~<~-0.04$  & $0.316~\pm~0.003$  \\ \cline{2-3} 
	&$-0.04~\leq~x_F~<~~0.00$  & $0.256~\pm~0.002$  \\ \cline{2-3}
	&$~0.00~\leq~x_F~<~+0.04$  & $0.197~\pm~0.002$  \\ \cline{2-3}
$\pi^-$	&$+0.04~\leq~x_F~<~+0.08$  & $0.146~\pm~0.002$  \\ \cline{2-3}
	&$+0.08~\leq~x_F~<~+0.12$  & $0.105~\pm~0.002$  \\ \cline{2-3}
	&$+0.12~\leq~x_F~<~+0.16$  & $0.085~\pm~0.003$  \\ \cline{2-3} 
	&$+0.16~\leq~x_F~<~+0.20$  & $0.083~\pm~0.005$  \\ \cline{2-3}
	&$+0.20~\leq~x_F~<~+0.30$  & $0.081~\pm~0.005$  \\ \cline{2-3}
	&$+0.30~\leq~x_F~\leq~+0.40$  & $0.095~\pm~0.013$  \\ \hline\hline

	&$-0.16~\leq~x_F~<~-0.12$ & $0.458~\pm~0.027$ \\ \cline{2-3}
	&$-0.12~\leq~x_F~<~-0.08$ & $0.408~\pm~0.014$ \\ \cline{2-3}
	&$-0.08~\leq~x_F~<~-0.04$ & $0.395~\pm~0.007$ \\ \cline{2-3} 
	&$-0.04~\leq~x_F~<~~0.00$ & $0.320~\pm~0.006$ \\ \cline{2-3}
	&$~0.00~\leq~x_F~<~+0.04$ & $0.314~\pm~0.005$ \\ \cline{2-3}
K$^-$	&$+0.04~\leq~x_F~<~+0.08$ & $0.312~\pm~0.006$ \\ \cline{2-3}
	&$+0.08~\leq~x_F~<~+0.12$ & $0.335~\pm~0.007$ \\ \cline{2-3}
	&$+0.12~\leq~x_F~<~+0.16$ & $0.402~\pm~0.009$ \\ \cline{2-3} 
	&$+0.16~\leq~x_F~<~+0.20$ & $0.479~\pm~0.012$ \\ \cline{2-3}
	&$+0.20~\leq~x_F~<~+0.30$ & $0.563~\pm~0.011$ \\ \cline{2-3}
	&$+0.30~\leq~x_F~\leq~+0.40$ & $0.749~\pm~0.025$ \\ \hline\hline
\end{tabular}
\protect\caption{$\Lm$ production asymmetries vs $x_F$ for negative beams. 
The errors shown are statistical, with systematic errors being negligible.}
\label{tabsysxfneg}
\end{table}
%%%
\begin{table} [ht] \centering
\begin{tabular} {|c|c|c|} \hline 
Beam	&${p_T}^2$ Region & Production Asymmetry    \\ \hline\hline

	&$0.00 <{{p_T}^2}\leq 0.10$ & $0.198~\pm~0.004$ \\ \cline{2-3} 
	&$0.10 <{{p_T}^2}\leq 0.26$ & $0.195~\pm~0.003$ \\ \cline{2-3}
	&$0.26 <{{p_T}^2}\leq 0.50$ & $0.196~\pm~0.003$ \\ \cline{2-3}
	&$0.50 <{{p_T}^2}\leq 0.80$ & $0.204~\pm~0.003$ \\ \cline{2-3}
	&$0.80 <{{p_T}^2}\leq 1.20$ & $0.225~\pm~0.004$ \\ \cline{2-3}
$\pi^-$	&$1.20 <{{p_T}^2}\leq 2.00$ & $0.245~\pm~0.005$ \\ \cline{2-3} 
	&$2.00 <{{p_T}^2}\leq 3.00$ & $0.266~\pm~0.008$ \\ \cline{2-3}
	&$3.00 <{{p_T}^2}\leq 4.00$ & $0.306~\pm~0.014$ \\ \cline{2-3}
	&$4.00 <{{p_T}^2}\leq 6.00$ & $0.302~\pm~0.022$ \\ \cline{2-3}
	&$6.00 <{{p_T}^2}\leq 10.0$ & $0.356~\pm~0.048$ \\ \hline\hline

	&$0.00 <{{p_T}^2}\leq 0.10$ & $0.336~\pm~0.012$ \\ \cline{2-3} 
	&$0.10 <{{p_T}^2}\leq 0.26$ & $0.317~\pm~0.009$ \\ \cline{2-3}
	&$0.26 <{{p_T}^2}\leq 0.50$ & $0.352~\pm~0.008$ \\ \cline{2-3}
	&$0.50 <{{p_T}^2}\leq 0.80$ & $0.388~\pm~0.009$ \\ \cline{2-3}
	&$0.80 <{{p_T}^2}\leq 1.20$ & $0.404~\pm~0.010$ \\ \cline{2-3}
K$^-$	&$1.20 <{{p_T}^2}\leq 2.00$ & $0.406~\pm~0.012$ \\ \cline{2-3} 
	&$2.00 <{{p_T}^2}\leq 3.00$ & $0.436~\pm~0.019$ \\ \cline{2-3}
	&$3.00 <{{p_T}^2}\leq 4.00$ & $0.459~\pm~0.033$ \\ \cline{2-3}
	&$4.00 <{{p_T}^2}\leq 6.00$ & $0.592~\pm~0.055$ \\ \cline{2-3}
	&$6.00 <{{p_T}^2}\leq 10.0$ & $0.540~\pm~0.160$ \\ \hline\hline
\end{tabular}
\protect\caption{$\Lm$ production asymmetries vs ${p_T}^2$ for negative beams. 
The errors shown are statistical, with systematic errors being negligible.}
\label{tabsysptneg}
\end{table}
%%%

\begin{thebibliography}{90} 

\bibitem{javier}
E. M. Aitala et al.(E791 Collaboration), \PL{B 496} (2000) 9.
\bibitem{charm-exp} 
E.M. Aitala et al. (E791 Collaboration), \PL{B 371} (1996) 157.
\bibitem{charm-expb}
G.A. Alves et al. (E769 Collaboration), \PRL{72} (1994) 812 and
{\bf 72} (1994) 1946.
\bibitem{charm-expc} 
M. Adamovich et al. (WA92 Collaboration), \NP{B 495} (1997) 3. 

\bibitem{sudeshna} E.M. Aitala et al. (E791 Collaboration), \PL{B 411} (1997)
230.
\bibitem{strange-modl} L. G. Pondrom, Phys. Rept. 122 (1985) 57.

\bibitem{charm-mod} V.G. Kartvelishvili, A.K. Likhoded, and S.R. 
Slobospitskii, Sov. J. Nucl. Phys. {\bf 33} (1981) 434.
\bibitem{charm-modb}  
R.C. Hwa, Phys. Rev. {\bf D 51} (1995) 85.
\bibitem{charm-modc} 
R. Vogt and S.J. Brodsky, Nucl. Phys. {\bf B 478} (1996) 311.
\bibitem{charm-modd} 
B.W. Harris, J. Smith, and R. Vogt, Nucl. Phys. {\bf B 461} (1996) 181.
\bibitem{charm-mode}
G. Herrera and J. Magnin, Eur. Phys. J. {\bf C 2} (1998) 477. 
%; E. Cuautle, G. Herrera and J. Magnin, Eur. Phys. J. {\bf C 2} (1998) 473. 

\bibitem{capella1} A. Capella, U. Sukhatme, C.I. Tan, and J. Tran  
Thanh Van, \PR{D 36} (1987) 109. 
 

\bibitem{accmor1} S. Barlag et al. (ACCMOR Collaboration), \PL{B 325} (1994) 531.

\bibitem{add-evid} S. Mikocki et al. (E580 Collaboration), \PR{D 34} (1986) 42.

\bibitem{add-evidb} R.T. Edwards et al. (E415 Collaboration), \PR{D 18} (1978) 76.

\bibitem{add-evidc} N.N. Biswas et al. (E002, E281, and E597 Collaborations), \NP{B 167} (1980) 41.

\bibitem{add-evidd} D. Bogert et al. (E234 Collaboration), \PR{D 16} (1977) 2098.

\bibitem{DISC} M. Benot, J. Litt, and R. Meunier, \NIM{105} (1972) 431.

\bibitem{TRD} D. Errede et al., \NIM Phys. Res., Sect {\bf A 309} (1991) 386.

\bibitem{jeff} J. Appel, \REV{42}  (1992) 367.

\bibitem{colin} C. Gay and S. Bracker, IEEE Trans. Nucl. Sci. {\bf 34} (1987)
870.

\bibitem{E769}  G. A. Alves et al. (E769 Collaboration), \PRL{69} (1992) 3147.

\bibitem{E769b} G.A. Alves et al. (E769 Collaboration), \PRL{77} (1996) 2388;
G.A. Alves et al. (E769 Collaboration), \PRL{81} (1998) 1537.


%\bibitem{E769c} G.A. Alves et al. (E769 Collaboration), \PRL{81} (1998) 1537.

\bibitem{wa82} M. Adamovich et al. (WA82 Collaboration), \PL{B 305} (1993) 402. 

\bibitem{pyth} T. Sj\"{o}strand, \PYTHIA ~5.7 and \JETSET ~7.4. Physics and 
Manual, CERN-TH-7112/93 (1993) hep-ph/9508391.

\bibitem{pythb} H.U. Bengtsson and T. Sj\"{o}strand, Comput. Phys. Commun. {\bf 46} (1987) 43.

%\bibitem{pythc} T. Sjostrand, CERN-TH.7112/93 (1993). 

\bibitem{pyth62} T. Sj\"ostrand, L. L\"onnblad, S. Mrenna, ``PYTHIA 6.2",
hep-ph/0108264, (2001). 

\bibitem{accmor2} S. Barlag et al. (ACCMOR Collaboration), \PL{B 233} (1989) 522. 


\bibitem{strange-mod} J.C. Anjos, J. Magnin, F.R.A. Sim\~ao, and J. Solano, 
Proceedings of II Silafae, AIP Conf. Proc. 444 (1998) 540, hep-ph/9806396.

%\bibitem{strange-mod2} W.G.D. Dharmaratna and G.R. Goldstein, \PR{D 41} (1990) 1731. 

%%\bibitem{hyp99} J.C. Anjos for the E791 Collaboration, Proceedings of the
%%Hyperon'99 Conference, Fermilab (27-29 September, 1999), hep-ex/991203.
%\bibitem{hq98} (E791 Collaboration), Proceedings of the Workshop on Heavy
%Quarks at Fixed Target, Fermilab, U.S.A. (10 October, 1998).

\end{thebibliography}
\end{document}